# Risk-based Classical Failure Mode and Effect Analysis (FMEA) of Microgrid Cyber-physical Energy Systems

Shravan Kumar Akula
School of Electrical Engineering
and Computer Sciences
University of North Dakota
Grand Forks, ND
shravankumar.akula@und.edu

Hossein Salehfar
School of Electrical Engineering
and Computer Sciences
University of North Dakota
Grand Forks, ND
h.salehfar@und.edu

*Abstract*—Modern microgrids are networked systems comprising physical and cyber components for networking, computation, and monitoring. These cyber components make microgrids more reliable but increase the system complexity. Therefore, risk assessment methods are an imperative technology for cyber-physical systems to ensure safe and secure operations. The authors propose a reliability approach that utilizes the failure modes of power and cyber network key components to perform the risk analysis in microgrids. In this work, the authors have used the Failure Mode and Effect Analysis (FMEA) approach for risk assessment of microgrid systems and determine the influence of various failure modes on their performance. FMEA is one of the most effective methods to assess the risk involving the cyber components. It is the method of examining components, modules, and subsystems to determine failure modes in a system and their causes and consequences. A novel approach is proposed to calculate risk priority numbers based on factors like severity, occurrence, and detection. A risk matrix is calculated from the proposed FMEA worksheet, which acts as a graphical representation for evaluating components risk based on risk priority number or criticality metrics.

## I. Introduction

The Department of Energy defines a microgrid (MG) as a group of interconnected loads and distributed energy resources within clearly defined electrical boundaries that acts as a single controllable entity with respect to the grid. An MG can connect and disconnect from the grid to enable it to operate in both grid-connected or island- mode [1]. Distributed generators in MG are renewable in nature. The intermittent nature of renewable energy sources is challenging as it can cause power fluctuations over multiple time horizons forcing the MG and utility grid operators to change the real-time operating procedures. Failure of the component or subsystem in the MG can lead to a cascade of failures and, in extreme cases, lead to a blackout.

Modern MGs use an array of complex digital communications technologies for a variety of operational uses and data measurements. Unfortunately, these cyber-physical components of the MG are prone to cyberattacks which can disable, damage, manipulate or steal the data from MG operators.

When such a fragile system is involved, there is a necessity for reliability quantification using relevant methodology and tools. To detect and eradicate the potential failures, the system must be analyzed at the component level, thereby improving the revenue and operational life of the components. The authors of this work have performed a comprehensive reliability analysis for MG in grid-tied mode [2] using fault tree analysis then calculated time-based metrics and importance measures which will further increase the reliability of the MG. In [2], cyber components are not included as a part of the reliability analysis as cyberattacks are not time-based failures. To address this issue and as a part of the reliability-centered maintenance evaluation for the MG, authors have used the FMEA to identify the potential cyber-attacks and their causes and effects on the MG.

This paper serves the following objectives: a. To present a systematic methodology of FMEA application for cyber-physical components in the MG environment and to demonstrate the applicability of this technique to this system, b. Present the results in the form of an FMEA report and criticality matrix to evaluate the risk level.

The following section presents a brief introduction to FMEA and a review of the applications of FMEA in MG and smart grids environments. This is followed by a FMEA analysis of MG's cyber-physical components, results, and conclusions and future work.

## II. Failure Modes and Effective Analysis (FMEA)

### A. Main concept

The primary goal of FMEA is to identify possible failure modes, analyze the causes and effects of various component failure modes and decide what may be done to prevent or decrease the likelihood of high-risk failures. The study's findings can assist risk analysts in identifying and correcting failure modes that negatively impact the system's performance during the design and development stages.

Criticality analysis in FMEA is performed using three risk factors, namely severity (S), occurrence (O), and detection (D). Using these three factors, a risk priority number (RPN) to prioritize the failure modes is computed as follows:

$$RPN = S * O * D$$

Each of the three factors is rated on a scale ranging from 1 to 10. A typical ranking system for the three risk factors is provided in tables I-III [3]. The greater the risk for product/system reliability, the higher the RPN of a failure mode. The failure modes will be ranked based on the RPN scores, and appropriate remedial/mitigation actions should be taken sequentially from the high-risk to low-risk failure modes.

TABLE I. TRADITIONAL RATINGS FOR SEVERITY (S) IN CLASSICAL FMEA [3]

| Rating | Severity (S) | Severity of Effect |
|---|---|---|
| 10 | Hazardous without warning | Highest severity ranking of a failure mode, occurring without warning and the consequence is hazardous |
| 9 | Hazardous with warning | Higher severity ranking of a failure mode, occurring with warning and the consequence is hazardous |
| 8 | Very high | Operation of system or product is broken down without compromising safe |
| 7 | High | Operation of system or product may be continued, but performance of system or product is affected |
| 6 | Moderate | Operation of system or product is continued, and performance of system or product is degraded |
| 5 | Low | Performance of system or product is affected seriously, and the maintenance is needed |
| 4 | Very low | Performance of system or product is less affected, and the maintenance may not be needed |
| 3 | Minor | System performance and satisfaction with minor effect |
| 2 | Very minor | System performance and satisfaction with slight effect |
| 1 | None | No effect |

TABLE II. TRADITIONAL RATINGS FOR OCCURRENCE (O) IN CLASSICAL FMEA [3]

| Rating | Occurrence (O) | Probable failure rate |
|---|---|---|
| 10 | Extremely high (inevitable failure) | ≥1 in 2 |
| 9 | Very high | 1 in 3 |
| 8 | Repeated failures | 1 in 8 |
| 7 | High | 1 in 20 |
| 6 | Moderately high | 1 in 80 |
| 5 | Moderate | 1 in 400 |
| 4 | Relatively low | 1 in 2000 |
| 3 | Low | 1 in 15,000 |
| 2 | Remote | 1 in 150,000 |
| 1 | Nearly impossible | ≤1 in 1,500,000 |

TABLE III. TRADITIONAL RATINGS FOR DETECTION (D) IN CLASSICAL FMEA [3]

| Rating | Detection (D) | Criteria |
|---|---|---|
| 10 | Absolutely impossible | Design control does not detect a potential cause of failure or subsequent failure mode or there is no design control |
| 9 | Very remote | Very remote chance the design control will detect a potential cause of the failure or subsequent failure mode |
| 8 | Remote | Remote chance the design control will detect a potential cause of failure or subsequent failure mode |
| 7 | Very low | Meager chance the design control will detect a potential cause of failure or subsequent failure mode |
| 6 | Low | Low chance the design control will detect a potential cause of failure or subsequent failure mode |
| 5 | Moderate | Moderate chance the design control will detect a potential cause of failure or subsequent failure mode |
| 4 | Moderately high | Moderately high chance the design control will detect a potential cause of the failure or subsequent failure mode |
| 3 | High | High chance the design control will detect a potential cause of failure or subsequent failure mode |
| 2 | Very high | Very high chance the design control will detect a potential cause of failure or subsequent failure mode |
| 1 | Almost certain | Design control will almost certainly detect a potential cause of failure or subsequent failure mode |

While FMEA is an effective method to perform risk analysis and preventive management, there are some drawbacks:

- Risk factors S, O, and D are treated with the same weight while omitting their relative importance.
- Different combinations of S, O, and D can produce the exact value of RPN. This could result in ignoring some high-risk failure modes.
- Classical FMEA only considers risk factors regarding safety, ignoring critical factors like economic impacts.

B. Procedure

A methodical approach should be followed to carry out an FMEA effectively. The general technique for performing a classical FMEA can be broken down into numerous steps as shown in Fig. 1.

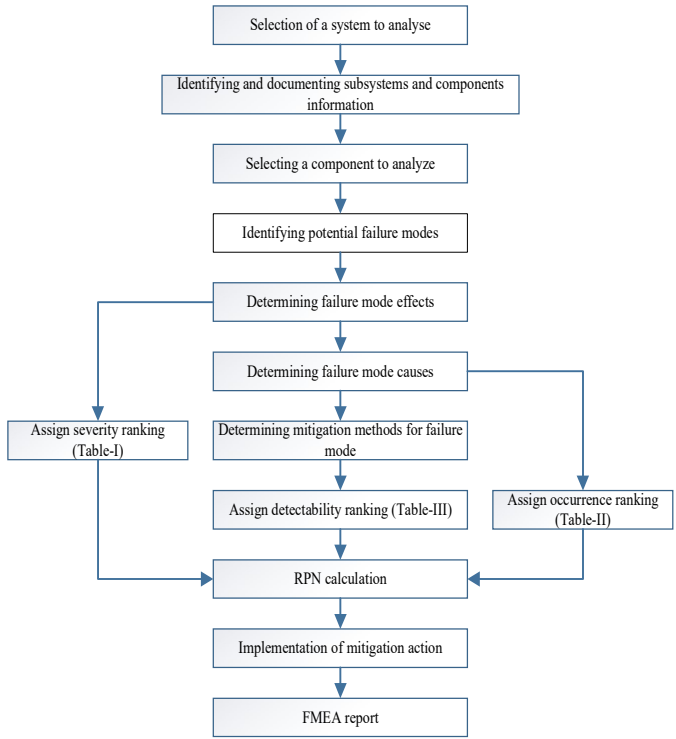

Fig.1. FMEA flowchart

### C. Literature review

FMEA has been proven to be a popular risk analysis technique and is multidisciplinary. Some work went into applying FMEA for cybersecurity. In [4] FMEA is used for assessing the likelihood of cybersecurity risks in manufacturing systems. Arben Asllani et al. [5] employs cybersecurity FMEA (C-FMEA) to show how failures of critical cyber components can affect the performance of the system by studying a regional airport as an example. The only work published for the application of FMEA in the context of smart grid cyber-physical systems is presented in [6] where failure modes and their effects on power and cyber components are discussed. In [6] a single FMEA risk analysis is performed for both cyber and power components, while for an in-depth and efficient analysis, they must be studied separately. To address this issue, authors of the present work are proposing an in-depth FMEA analysis considering only MG cyber components.

Following are some of the popular applications of FMEA in renewable energy sources.

- Photovoltaic systems: Colli Alessandra et al. [7] present an FMEA analysis for photovoltaic system components and sub-components. In [8] reliability-centered maintenance is carried out using FMEA and real data derived from a photovoltaic farm is used for the calculation of RPN.
- Wind turbines: In [9], an FMEA analysis is used to study the 2MW wind turbine and their assemblies. This work identifies and documents potential failure modes for key assemblies of a wind turbine and how their failure affects the overall performance. To reduce the total failure cost in wind turbines, a quantitative approach of FMEA using cost sensitivity analysis is proposed in [10].

## III. FMEA ANALYSIS FOR MG CYBER-PHYSICAL COMPONENTS

### A. Description of cyber components in MG

The safe and stable operation of MG requires the coordination of its cyber systems. The physical part of the MG contains distributed generators, primarily renewable energy sources, battery energy storage systems, loads, and other equipment. The cyber part contains data acquisition devices, controllers, state detectors, communication devices, and other equipment.

In [11] MG, a cybersecurity architecture is proposed to mitigate potential cyberattacks, and all the typical cyber components are listed. The list of cyber-physical components used in this study and their role in MG communication infrastructure is as follows : Energy management systems (EMS) to optimize MG operations, to transmit and receive data from physical equipment, server for storing MG operational data, Human-machine interface (HMI) for interaction with MG operators, intelligent electronic devices (IED) for sending data from relays, controllers to EMS, protection relays, generator and renewable energy controller for monitoring the controller data, load controller for sending load data to EMS, smart meter to collect data about energy transfer to and from the MG, remote terminal units (RTU) for transmitting field devices data to EMS, phasor measurement units (PMU) to transmit phasor data to EMS, battery energy storage controller for transmitting and receiving data from EMS to improve the power quality and to optimize the charging and discharging schedules, plug-in hybrid electric vehicle (PHEV) to interact with charging station, and software updates, and PHEV supply equipment to send charging/discharging information from charging station to EMS.

### B. Identifying potential failure modes for MG cyber components

Potential failure modes considered in this work are typical cyberattacks. A cyberattack is an attempt to gain unauthorized access to a system with an intent to damage, disable, or stealing the data. An overview of different types of cyber-attacks used in this work and their definitions are provided in [12] and [13]. Guneet et al. [14] documented different types of hacking and reasons to hack, data breaches and reasons for them, malicious attacks, outsider attacks, website vulnerabilities, different types of malware attacks, and the necessity for cyber laws. Authors in [11] documented vulnerabilities in internet protocol and industrial control systems networks.

Authors of the present work have documented different cyber-attacks from literature and used them in FMEA analysis.

### C. FMEA analysis and results

Using the MG cyber components and different types of cyber-attacks associated with the previous sections, a complete FMEA analysis for MG cyber-physical components was performed, and results are documented. Not only did authors perform a classical FMEA analysis, but they have also documented prevention controls, detection controls, and classification (based on RPN score) for each cyber component.

Ratings for risk factors are assigned according to standard FMEA evaluator's criteria. However, it is to be noted that risk factor ratings depend on the events that trigger them. In general, various Detection and Occurrence ratings are predicted for each failure mode, depending on the circumstances that prompted them, but the Severity rating for each failure mode is distinctive. In some cases, this can lead to different RPN ratings for the same failure mode. In this study, a total of fifteen cyber components with high-risk failure modes are identified and analyzed.

Due to the page restriction, authors are only able to include the complete FMEA report for top three critical components with highest RPN in Table V. For the rest of components risk factor scores, RPN rating, and classification are documented in Table IV. Detailed information and results from this work are directly available from the authors and will be included in other future works.

TABLE IV. RPN SCORE FOR REST OF CYBER COMPONENTS

| MG Cyber-Physical component | S | O | D | RPN | Classification |
|---|---|---|---|---|---|
| Database | 4 | 6 | 4 | 96 | Marginal |
| Server | 7 | 6 | 4 | 168 | Marginal |
| Intelligent electronic device (IED) | 7 | 5 | 4 | 140 | Critical |
| Generator controller | 6 | 5 | 4 | 120 | Critical |
| Automatic transfer switch (ATS) | 5 | 4 | 4 | 80 | Marginal |
| Renewable energy controller | 6 | 5 | 4 | 120 | Marginal |
| Remote terminal unit (RTU) | 5 | 7 | 6 | 210 | Critical |
| Phasor measurement unit (PMU) | 5 | 4 | 6 | 120 | Marginal |
| Disconnect switch | 7 | 4 | 4 | 112 | Marginal |
| PHEV | 9 | 5 | 4 | 180 | Catastrophic |
| PHEV supply equipment | 8 | 5 | 4 | 160 | Critical |
| Relay | 8 | 6 | 4 | 192 | Marginal |

It is necessary to account for some information loss during a traditional FMEA procedure to obtain FMEA results. As a consequence, essential findings about high-risk failure modes and their impact on system reliability may be jeopardized because of this predicament. This FMEA analysis coincides with the results in [6], proving the effectiveness of the proposed method.

A risk matrix provides a visual way for MG operators to assess risk based on the RPN rating. Figs. 2 and 3 provide the risk matrix for the proposed FMEA approach.

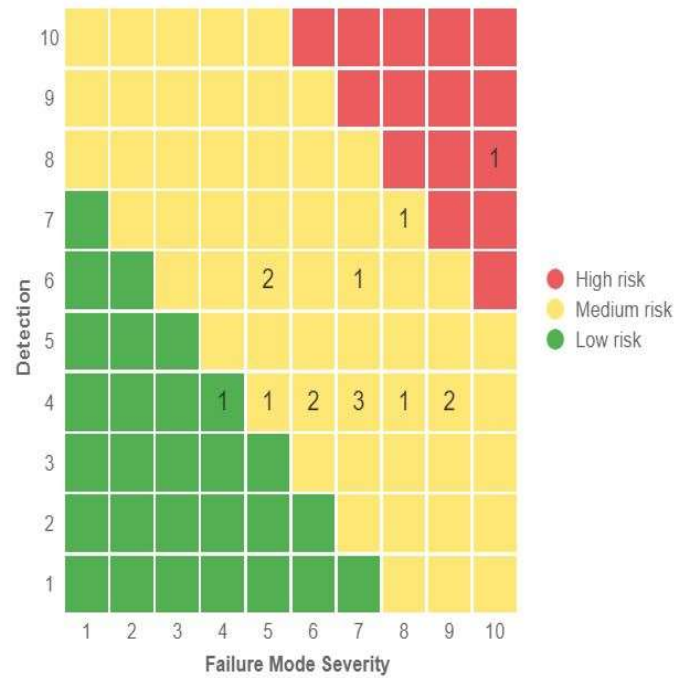

Fig. 2. Risk matrix (Severity vs Detection)

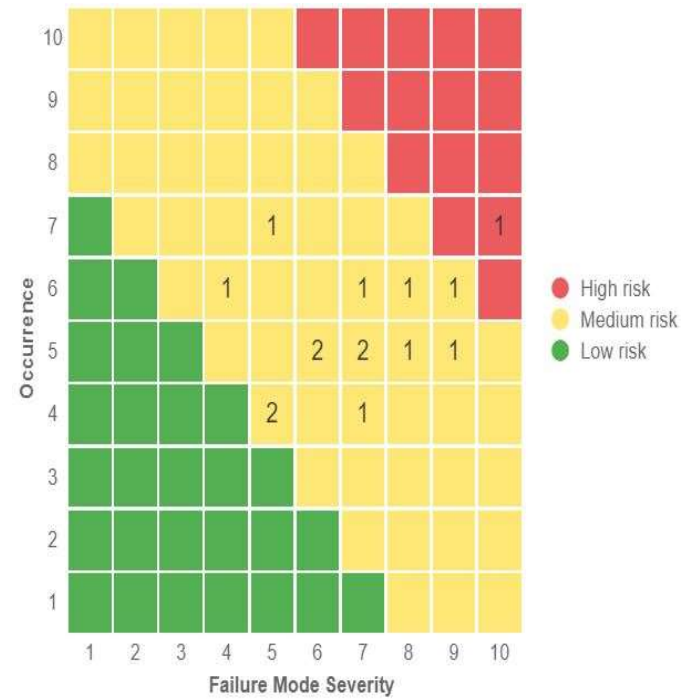

Fig. 3. Risk matrix (Severity vs Occurrence)

To find the cyber components corresponding to an element in the risk matrix, refer to the risk factors rating from the given FMEA report.

TABLE V.  FMEA REPORT FOR TOP THREE CRITICAL COMPONENTS

| MG Cyber-Physical component | Failure Mode | S | Effect | End effect | Cause | Classification | O | Prevention Controls | Detection Controls | D | RPN |
|---|---|---|---|---|---|---|---|---|---|---|---|
| Energy Management System (EMS) | While EMS is not a component, failure of one or more critical components listed below can lead to the failure of the EMS. Along with the cyberattacks listed below, some of the other factors that can lead to the EMS failure are listed here: insecure inter-network communication channel, default accounts on the components, default and weak passwords, weak authentication, weak coding practices, man-in-the-middle attack, weak/ misconfigured firewalls, exploitable ports, social engineering, and data buffer overflow because of the unchecked data stream | 10 | MG islanding without the grid operators' consent, false tripping of relays/ circuit breakers, overloading on the generation and distribution system, voltage/frequency violations, failed system protection, and unstable system | Hefty fines imposed by the government because of voltage/ frequency violations, loss of power to critical loads, grid instability, and cascading failures leading to MG failure | Unauthorized access to key components, lack of penetration testing, weak firewall/ antivirus, weak encryption, and lack of multifactor authentication | Catastrophic | 7 | Strict vulnerability and incident handling protocols, secured system architecture, removing unnecessary software, blocking unused ports, hardened account management, and access control policies, appropriate data backup and restore measures, intrusion detection systems, regular security patches, and updates, strong encryption techniques, strong passwords, using proxy servers, and strict file system access policies | Smart inverters, smart meters, comparing with historical SCADA measurements, intrusion detection software, real-time intrusion detection based on advanced machine learning algorithms, and key data sources that could predict potential attacks | 8 | 560 |
| Human-machine interface (HMI) | Hackers tend to target the HMI as the MG operators use this to interact with the SCADA system. Typical attacks on the HMI include malware, Stuxnet, insider threat, sniffer attack, man-in-the-middle attack, code injection, memory corruption, insecure default settings, plain text SCADA protocol, key stroking software, and network flooding | 8 | Unauthorized access to critical components by which hacker can change the settings (opening and closing of circuit breakers), permanent damage to the components, and loss of critical information | Violation of regulations imposed by governments can lead to hefty fines, reputational damage to MG owners and systems offline | Weak passwords, lack of relevant knowledge, default system settings, external device connected to network, suspicious emails, using the infected external storage device, and malware | Critical | 6 | Segregation and increased perimeter security, using honeypots to find vulnerabilities in the system, advanced firewall installation, intrusion detection systems, demilitarized zone to increase security at MG site, using virtual private networks, and encryption | Periodical risk assessment, intrusion detection software installation, MAC address locking, risk auditing, using machine learning algorithms, strong passwords, and relevant personnel training | 7 | 336 |
| Smart meter | Power denial attack, power theft attack, grid disruption attack, internet protocol misconfiguration, memory corruption, promotion request overflow, node log overflow, identity theft, overflow in the local switch identifier, intercepting traffic, central processing unit overload, traffic reinjection, physical layer jamming, access control jamming, physical attack | 7 | Power outage to customers and critical loads, power theft from the utility, instability in the grid, widespread power loss, data loss, and financial loss to MG and utility grid operators | Power outage and monetary loss | Jamming the communication channel, control media access jamming, vulnerabilities in the firewall, complex network topology, older switches, outdated firmware, and software | Marginal | 5 | Electromagnetic field radiation shield, Strong firewall, command source authentication, encryption, load drop detection, and implementing head-end access controls | Implementing signature-based and anomaly-based intrusion detection software | 6 | 210 |

IV. Conclusions and Future work

This paper discusses the most critical failure modes for the cyber-physical components and analyzes the application of classical FMEA in the MG environment. First, the results of a qualitative reliability study were completed along with a critical review of FMEA outcomes. Once the failure modes are identified, FMEA analysis was conducted, and an FMEA report was generated. This analysis will help the MG operators and designers to identify the weak points in cyber systems infrastructure. The proposed FMEA analysis has the potential to improve the reliability of the MG cyber-physical system, where reliability plays a crucial role in cost-effectiveness.

Future research in FMEA for cyber-physical components and MG physical components holds a lot of promise. To address gaps in this study, authors will include the economic risk assessment of the MG components, and FMEA analysis will be expanded to MG physical components. Also, calculated RPN will be compared with field reliability data, and similarities among them will be documented.

In classical FMEA, for RPN calculation, the degree of importance for risk factors is not considered. To address this issue, fuzzy-FMEA will be used in future work.